\documentclass[aps,preprint]{revtex4}
\usepackage[dvips]{graphicx}

\makeatletter
\@addtoreset{equation}{section}
\makeatother

\begin{document}

\preprint{KOBE-TH-06-01}

\title{
Relation between CKM and MNS Matrices \\Induced by Bi-Maximal Rotations in the Seesaw Mechanism
}


\author{H.~Tsujimoto}
\email[]{tujimoto@kobe-u.ac.jp}
\affiliation
{Graduate School of Science and Technology, Kobe University, \\Rokkodaicho 1-1, Nada ward, Kobe 657-8501}



\begin{abstract}
It is found that the seesaw mechanism not only explains the smallness of neutrino masses but also accounts for the large mixing angles simultaneously, even if the unification of the neutrino Dirac mass matrix with that of up-type quark sector is realized.
In this mechanism, we show that the mixing matrix of the Dirac-type mass matrix gets extra rotations from the diagonalization of Majorana mass matrix.
Assuming that the mixing angles to diagonalize the Majorana mass matrix are extremely small, we find that the large mixing angles of leptonic sector found in atmospheric and long baseline reactor neutrino oscillation experiments can be explained by these extra rotations.
We also find that provided the mixing angle around y-axis to diagonalize the Majorana mass matrix vanishes, we can derive the information about the absolute values of neutrino masses and Majorana mass responsible for the neutrinoless double beta decay experiment through the data set of neutrino experiments.
In the simplified case that there is no CP phase, we find that the neutrino masses are decided as $m_1:m_2:m_3\approx 1:2:8$ and that there are no solution which satisfy $m_3<m_1<m_2$ (inverted mass spectrum).
Then, including all CP phases, we reanalyze the absolute values of neutrino masses and Majorana mass responsible for the neutrinoless double beta decay experiment.
\end{abstract}



\maketitle

\section{Introduction\label{intro}}
Neutrino sector has many curious properties which are not shared by the quark and charged leptonic sectors.
For example, neutrino masses are very small \cite{WMAP} compared with those of quarks and charge leptons.
The large mixing angles seen in the experiments of atmospheric neutrino oscillation and long baseline reactor neutrino oscillation experiments (related to solar neutrino deficit) \cite{SK1,SK2,SK3,Kam1,Kam2,SNO1,SNO2} are also new features, not seen in the quark sector.

It is well known that the seesaw mechanism \cite{Yanagida,Gell-Mann,Mohapatra} can explain the small mass scale of neutrinos naturally.
In this mechanism, neutrino mass matrix which describes the low energy observables is given approximately by
\begin{eqnarray}
\mathcal{M}_{\nu}
=
-(\mathcal{M}_D)^T(\mathcal{M}_R)^{-1}(\mathcal{M}_D),
\label{numass1}
\end{eqnarray}
where $\mathcal{M}_D$ and $\mathcal{M}_R$ are the Dirac and Majorana mass matrices of neutrino, respectively.
The unpleasant overall minus sign can be absorbed by redefinition of field as $\nu\rightarrow i\gamma^5\nu$, i.e.
\begin{eqnarray*}
\mathcal{L}_{mass}
&=&
-\overline{\left(\nu _L\right)^c}\mathcal{M}_{\nu}\nu _L+h.c.
=
-\nu^TC\left(\frac{1-\gamma^5}{2}\right)\mathcal{M}_{\nu}\nu+h.c.
\\
&\rightarrow&
-\nu^TC\left(\frac{1-\gamma^5}{2}\right)(-\mathcal{M}_{\nu})\nu+h.c.
\end{eqnarray*}
If we require that the order of magnitude of $\mathcal{M}_D$ is the weak scale and that of $\mathcal{M}_R$ is the GUT scale, we can roughly obtain the desired order of magnitude of $\mathcal{M}_{\nu}$.

In addition, this mechanism can also explain the large mixing angles in the leptonic sector inheriting the unification of lepton and quark sectors as is seen in $SO(10)$ GUT.
Especially, it has been pointed out in some articles (e.g. \cite{Smirnov}) that there exist interesting and amusing relations between CKM and MNS matrices :
\begin{eqnarray}
&\bullet&\theta _{sol}+\theta _{Cabibbo}\simeq 45^{\circ}
\label{rel1}
\\
&\bullet&\theta _{atm}+\theta _{23}^{CKM} \simeq 45^{\circ}
.
\label{rel2}
\end{eqnarray}
These relations may imply that there exist some nontrivial relations between CKM and MNS matrices and that the seesaw mechanism has a comparatively simple structures as seen below.

To clarify our procedures, we use the following notations, i.e. $\mathcal{M}_D$ and $\mathcal{M}_R$ are diagonalized as
\begin{eqnarray*}
\mathcal{V}_R^{\dag}\mathcal{M}_D\mathcal{V}_L=\hat{\mathcal{M}}_D
\qquad
\mathcal{U}^T\mathcal{M}_R\mathcal{U}=\hat{\mathcal{M}}_R,
\end{eqnarray*}
where $\hat{\mathcal{M}}_D$ and $\hat{\mathcal{M}}_R$ are diagonalized Dirac and Majorana mass matrices, respectively and $\mathcal{V}_{L,R}$ and $\mathcal{U}$ are unitary matrices.
Using these notations in Eq.(\ref{numass1}), $\mathcal{M}_{\nu}$ can be written as
\begin{eqnarray}
\mathcal{M}_{\nu}
&=&
\mathcal{V}_L^{\ast}(\hat{\mathcal{M}}_D)\mathcal{V}_R^T\cdot
\mathcal{U}(\hat{\mathcal{M}}_R)^{-1}\mathcal{U}^T\cdot
\mathcal{V}_R(\hat{\mathcal{M}}_D)\mathcal{V}_L^{\dag}
\nonumber\\
&=&
\mathcal{V}_L^{\ast}(\hat{\mathcal{M}}_D)U_R(\hat{\mathcal{M}}_R)^{-1}U_R^T(\hat{\mathcal{M}}_D)\mathcal{V}_L^{\dag}
,
\label{numass2}
\end{eqnarray}
where we define a unitary matrix, $U_R=\mathcal{V}_R^T\mathcal{U}$.
In $SO(10)$ GUT, there are some nontrivial relations between quark and leptonic sectors and furthermore between $\mathcal{V}_L$ and $\mathcal{V}_R$ above the symmetry breaking scale, which we adopt in this work.
This is because $SO(10)$ includes a subgroup, $G=SU(4)_{PS}\times SU(2)_L\times SU(2)_R$.
$SU(4)_{PS}$ symmetry leads to the relations between quark and lepton Yukawa coupling matrices, i.e.
\begin{eqnarray}
Y_u=Y_{\nu}
\qquad
Y_d=Y_e
,
\label{su4}
\end{eqnarray}
where the indices of $u,d,\nu,e$ correspond to up-type quark, down-type quark, neutrino, charged-lepton, respectively.
In addition, $SU(2)_L\times SU(2)_R$ symmetry leads to left-right symmetry.
Since the two indices to denote the matrix elements of Dirac-type mass matrix correspond to left- and right-handed neutrinos, this symmetry reduces the degrees of freedom of the matrix, i.e. it should be a symmetric matrix, and this leads to a relation
\begin{eqnarray}
\mathcal{V}_R=\mathcal{V}_L^{\ast}.
\label{su2}
\end{eqnarray}
For simplicity of the argument, we assume that these relations hold approximately at low energies.
Adopting a certain basis in which the down-type quark mass matrix is diagonalized, the former relation in Eq.(\ref{su4}) leads to
\begin{eqnarray}
\mathcal{V}_L^{\dag}=V_{CKM}.
\label{lckm}
\end{eqnarray}
Then, we can rewrite Eq.(\ref{numass2}) as
\begin{eqnarray}
\mathcal{M}_{\nu}
=
V_{CKM}^T(\hat{\mathcal{M}}_D)U_R(\hat{\mathcal{M}}_R)^{-1}U_R^T(\hat{\mathcal{M}}_D)V_{CKM}
.
\label{numass3}
\end{eqnarray}
The r.h.s. of Eq.(\ref{numass3}) is furthermore diagonalized as
\begin{eqnarray}
\mathcal{M}_{\nu}
=
V_{CKM}^TO^{\ast}(\hat{\mathcal{M}}_{\nu})O^{\dag}V_{CKM},
\label{numass4}
\end{eqnarray}
where $\hat{\mathcal{M}}_{\nu}$ is the diagonalized neutrino mass matrix with mass eigenvalues $\mu _1,\mu _2$ and $\mu _3$.
The matrix $V_{CKM}^{\dag}\times O$ is what we call MNS matrix of leptonic sector
\begin{eqnarray}
\left[
\begin{array}{c}
\nu _e\\
\nu _{\mu}\\
\nu _{\tau}\\
\end{array}
\right]
=
\left[
\begin{array}{ccc}
&&\\
&V_{MNS}&\\
&&\\
\end{array}
\right]
\left[
\begin{array}{c}
\nu _1\\
\nu _2\\
\nu _3\\
\end{array}
\right].
\label{mns}
\end{eqnarray}
In this way, the mixing matrix, $V_{CKM}$, is accompanied by extra rotations by $O$, so that we can explain the disagreement between CKM and MNS matrices and the large mixing angles of leptonic sector once $O$ contains large (maximal) mixing angles.

In this manuscript, we especially concentrate our attention on the relations found in Eqs.(\ref{rel1}),(\ref{rel2}).
As we sketch right below, these relations are realized provided the extra rotations due to $O$ are bi-maximal rotations around x- and z-axes.
This may be natural since these relations are concerning $1\leftrightarrow 2$ and $2\leftrightarrow 3$ generation mixings.

According to an approximation proposed by Wolfenstein \cite{Wolf}, we can parametrize $V_{CKM}$ as
\begin{eqnarray*}
V_{CKM}
=
\left[
\begin{array}{ccc}
1-\frac{\lambda^2}{2}&\lambda&A\lambda^3\rho(1-i\eta)\\
-\lambda&1-\frac{\lambda^2}{2}&A\lambda^2\\
A\lambda^3(1-\rho-i\eta)&-A\lambda^2&1\\
\end{array}
\right],
\end{eqnarray*}
where $\lambda\simeq\sin\theta _C$ and $A,\rho$ and $\eta$ are quantities of the order of unity.
Roughly speaking, since the bi-maximal extra rotations shifts these angles to
\begin{eqnarray*}
-\lambda
&\rightarrow&
-\lambda+45^{\circ},
\\
-A\lambda^2
&\rightarrow&
-A\lambda^2+45^{\circ},
\end{eqnarray*}
we can obtain the desired relations in Eq.(\ref{rel1}) and Eq.(\ref{rel2}).
\begin{figure}[h]
\begin{center}
\includegraphics[scale=1.0]{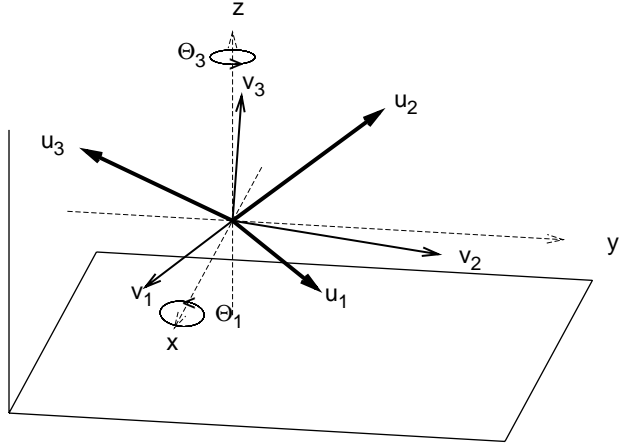}
\end{center}
\caption{A geometrical relation between CKM and MNS matrices.}
\label{figKMMNS}
\end{figure}

We can easily understand the relation between CKM and MNS matrices geometrically.
The explicit forms of $V_{CKM}$ and $V_{MNS}^{\dag}$ with a standard parametrization are given by
\begin{eqnarray*}
V_{CKM}
\approx
\left[
\begin{array}{ccc}
0.9745&0.2243&0.0037\\
-0.2243&0.9737&0.0413\\
0.0057&-0.0411&0.9991\\
\end{array}
\right]
\qquad
V_{MNS}^{\dag}
\approx
\left[
\begin{array}{ccc}
0.8482&-0.3746&0.3746\\
0.5297&0.5998&-0.5998\\
0&0.7071&0.7071\\
\end{array}
\right],
\end{eqnarray*}
where we set $\theta _{12}^{MNS}=\theta _{sol}=32^{\circ},\theta _{23}^{MNS}=\theta _{atm}=45^{\circ},\theta _{13}^{MNS}=\theta _{CHOOZ}=0^{\circ}$ and ignore CP phases tentatively.
Decomposing these matrices to vector representations
\begin{eqnarray*}
V_{CKM}
=
\left[
\begin{array}{ccc}
\vec{v_1}&\vec{v_2}&\vec{v_3}\\
\end{array}
\right]
\qquad
V_{MNS}^{\dag}
=
\left[
\begin{array}{ccc}
\vec{u_1}&\vec{u_2}&\vec{u_3}\\
\end{array}
\right],
\end{eqnarray*}
we can express these in vector space as shown in Fig.\ref{figKMMNS}.
We parametrize the orthogonal matrix $O$ as $O=O_x(\Theta _1)\cdot O_z(\Theta _3)$.
Then, from the relation between CKM and MNS matrices, $V_{MNS}^{\dag}=O^{\dag}\times V_{CKM}$, we can easily see that $\vec{v}_i$ is rotated around x-axis by $\Theta _1$ first and around z-axis by $\Theta _3$ next to get $\vec{u}_i$.
Thus, we can roughly achieve the above relations, once $\Theta _1$ and $\Theta _3$ are (almost) maximal.

In what follows, we parametrize the unitary matrices $U_R$ by three mixing angles and five CP phases after absorbing one overall CP phase by the rephasing of fields, i.e.
\begin{eqnarray}
U_R
&=&
P_1\times V_R\times P_2
\label{udef}
\\
V_R
&=&
\left[
\begin{array}{ccc}
1&0&0\\
0&\cos\theta _1^R&\sin\theta _1^R\\
0&-\sin\theta _1^R&\cos\theta _1^R\\
\end{array}
\right]
\left[
\begin{array}{ccc}
\cos\theta _2^R&0&e^{-i\delta^R}\sin\theta _2^R\\
0&1&0\\
-e^{i\delta^R}\sin\theta _2^R&0&\cos\theta _2^R\\
\end{array}
\right]
\left[
\begin{array}{ccc}
\cos\theta _3^R&\sin\theta _3^R&0\\
-\sin\theta _3^R&\cos\theta _3^R&0\\
0&0&1\\
\end{array}
\right]
\nonumber\\
P_i
&\stackrel{def}{=}&
diag\left(e^{i\epsilon _i},1,e^{i\kappa _i}\right),
\nonumber
\end{eqnarray}
where we use a standard parametrization of $V_R$ \cite{PDG,Chau,Harari,Fritzsch,Botella} in which three mixing angles, $\theta _1^R,\theta _2^R,\theta _3^R$, and one CP phase, $\delta^R$, are embedded.

Throughout this manuscript, we perform analyses regarding the three mixing angles to be small, i.e. $\theta _1^R,\theta _2^R,\theta _3^R\ll 1$.
The reason can be understood as follows.
For simplicity, let us first think about two generation case.
In this case, the mass matrix, $\mathcal{M}=(\hat{\mathcal{M}}_D)V_R(\hat{\mathcal{M}}_R)^{-1}V_R^T(\hat{\mathcal{M}}_D)$, can be explicitly written as
\begin{eqnarray*}
\mathcal{M}
&=&
\left[
\begin{array}{cc}
m_1&\\
&m_2\\
\end{array}
\right]
\left[
\begin{array}{cc}
\cos\theta^R&\sin\theta^R\\
-\sin\theta^R&\cos\theta^R\\
\end{array}
\right]
\left[
\begin{array}{cc}
\frac{1}{M_1}&\\
&\frac{1}{M_2}\\
\end{array}
\right]
\left[
\begin{array}{cc}
\cos\theta^R&-\sin\theta^R\\
\sin\theta^R&\cos\theta^R\\
\end{array}
\right]
\left[
\begin{array}{cc}
m_1&\\
&m_2\\
\end{array}
\right]
\\
&=&
(\cos\theta^R)^2\frac{m_1^2}{M_1}
\times
\left[
\begin{array}{cc}
1+(\tan\theta^R)^2\frac{M_1}{M_2}&-\tan\theta^R\left(1-\frac{M_1}{M_2}\right)\frac{m_2}{m_1}\\
\tan\theta^R\left(1-\frac{M_1}{M_2}\right)\frac{m_2}{m_1}&\left((\tan\theta^R)^2+\frac{M_1}{M_2}\right)\left(\frac{m_2}{m_1}\right)^2\\
\end{array}
\right]
,
\end{eqnarray*}
where $\hat{\mathcal{M}}_D=diag(m_1,m_2),\hat{\mathcal{M}}_R=diag(M_1,M_2)$ and we ignore all CP phases.
The necessary and sufficient condition for this mass matrix to be diagonalized by a orthogonal matrix with maximal rotation, i.e. $\mathcal{M}=O\cdot(\hat{\mathcal{M}})\cdot O^T$ and $O=O(45^{\circ})$, is $\mathcal{M}_{11}=\mathcal{M}_{22}$, which in turn leads to the following relation,
\begin{eqnarray*}
\left(\tan\theta^R\right)^2
=
\frac{\left(\frac{m_1}{m_2}\right)^2-\frac{M_1}{M_2}}{1-\frac{M_1}{M_2}\left(\frac{m_1}{m_2}\right)^2}
.
\end{eqnarray*}
We make a natural assumption that the Majorana mass eigenvalues have hierarchical structure, i.e. $M_1/M_2\ll 1$, while $\left(m_1/m_2\right)^2=\left(m_u/m_c\right)^2\ll 1$ as expected by the quark-lepton symmetry in $SO(10)$ GUT.
Thus, we can approximate the mixing angle as
\begin{eqnarray}
(\theta^R)^2
\simeq
\left(\frac{m_1}{m_2}\right)^2-\frac{M_1}{M_2}
\ll 1.
\label{}
\end{eqnarray}
Note that the requirement for maximal rotation is equivalent to the small mixing angle of Majorana mass matrix, once if we assume that the Majorana mass eigenvalues have hierarchical structure.
Furthermore, we can estimate the order of magnitude of the mixing angle as
\begin{eqnarray}
\theta^R
=
\mathcal{O}\left(\frac{m_1}{m_2}\right)
\quad
\textrm{or}
\quad
\mathcal{O}\left(\sqrt{\frac{M_1}{M_2}}\right)
.
\label{}
\end{eqnarray}
It is interesting to note that the latter relation, $\theta^R=\mathcal{O}\left(\sqrt{M_1/M_2}\right)$, is what we find in quark sector approximately (for Dirac masses, though).
Being Inspired by this discussion, in the realistic three generation case, we regard the three mixing angles, $\theta _1^R,\theta _2^R,\theta _3^R$, to be small.
Comparing with the data on neutrino oscillations, it turns out that the relations, $\theta _{ij}^R=\mathcal{O}\left(\sqrt{M_i/M_j}\right)$, hold well  in the three generation analysis.

This paper is organized as follows.
In Sec.\ref{two}, we discuss the extra rotations, sketched above, more carefully and emphasize that we can not only realize the relation (\ref{rel1}),(\ref{rel2}), but also derive the absolute values of neutrino masses, by comparing with the existing experimental data of neutrino oscillations.
The effects of five CP phases which embedded in $U_R$, i.e. $\epsilon _i,\kappa _i$ and $\delta^R$, are discussed in Sec.\ref{three}. 
\section{Bi-maximal extra rotations and estimation of absolute values of neutrino masses \label{two}}
In this section, we investigate neutrino mass matrix by switching off the five CP phases for simplicity.
Recalling Eq.(\ref{numass3}),
\begin{eqnarray}
\mathcal{M}_{\nu}
=
V_{CKM}^{\dag}(\hat{\mathcal{M}}_D)V_R(\hat{\mathcal{M}}_R)^{-1}V_R^T(\hat{\mathcal{M}}_D)V_{CKM}^{\ast},
\label{numass5}
\end{eqnarray}
we express the diagonalized mass matrices as
\begin{eqnarray*}
\hat{\mathcal{M}}_D=diag(m_1,m_2,m_3)
\qquad
\hat{\mathcal{M}}_R=diag(M_1,M_2,M_3)
\end{eqnarray*}
and parametrize the matrix $V_R$ in a specific form by two rotations around y- and z-axes as
\begin{eqnarray}
V_R
=
\left[
\begin{array}{ccc}
\cos\theta _2^R&0&\sin\theta _2^R\\
0&1&0\\
-\sin\theta _2^R&0&\cos\theta _2^R\\
\end{array}
\right]
\left[
\begin{array}{ccc}
\cos\theta _3^R&\sin\theta _3^R&0\\
-\sin\theta _3^R&\cos\theta _3^R&0\\
0&0&1\\
\end{array}
\right].
\label{vr}
\end{eqnarray}
In general, we should parametrize $V_R$ by three mixing angles as seen in Eq.(\ref{udef}).
We, however, can achieve bi-maximal rotations with a minimal set of mixing angles as is seen in Eq.(\ref{vr}) and this can be understood as follows.
Defining
\begin{eqnarray}
\mathcal{M}
&=&
\frac{M_1}{m_1^2}\times
(\hat{\mathcal{M}}_D)V_R(\hat{\mathcal{M}}_R)^{-1}V_R^T(\hat{\mathcal{M}}_D)
\nonumber\\
&=&
\left[
\begin{array}{ccc}
1&&\\
&\frac{m_2}{m_1}&\\
&&\frac{m_3}{m_1}\\
\end{array}
\right]
V_R
\left[
\begin{array}{ccc}
1&&\\
&\frac{M_1}{M_2}&\\
&&\frac{M_1}{M_3}\\
\end{array}
\right]
V_R^T
\left[
\begin{array}{ccc}
1&&\\
&\frac{m_2}{m_1}&\\
&&\frac{m_3}{m_1}\\
\end{array}
\right]
,
\label{mass1}
\end{eqnarray}
we can estimate diagonal terms as $\mathcal{M}_{ii}\sim M_1/M_i\left(m_i/m_1\right)^2$ and off-diagonal terms as
\begin{eqnarray*}
\mathcal{M}_{12}=\frac{m_2}{m_1}\theta _3^R
\qquad
\mathcal{M}_{13}=\frac{m_3}{m_1}\theta _2^R
\qquad
\mathcal{M}_{23}=\frac{m_2m_3}{m_1^2}\frac{M_1}{M_2}\theta _1^R
,
\end{eqnarray*}
where we assume that these angles are extremely small as mentioned in previous section.
Requiring that the order of magnitude of these quantities are unity, we find
\begin{eqnarray}
\frac{M_1}{M_i}=\mathcal{O}\left(\frac{m_1^2}{m_i^2}\right)
\qquad
\theta _1^R=\mathcal{O}\left(\frac{m_2}{m_3}\right)
\qquad
\theta _2^R=\mathcal{O}\left(\frac{m_1}{m_3}\right)
\qquad
\theta _3^R=\mathcal{O}\left(\frac{m_1}{m_2}\right)
.
\label{massrel}
\end{eqnarray}
Using these naive estimations, we can easily find that the choice of $V_R$ in Eq.(\ref{vr}) just leads to the conditions that $\mathcal{M}_{22}\simeq \mathcal{M}_{33}$ and $\mathcal{M}'_{22}\simeq \mathcal{M}_{11}$ (where $\mathcal{M}'_{22}$ denotes a matrix element after rotating by $\Theta _1$), once we expand the allowed region of $M_1$ to negative region ($\epsilon _2=0$ or $\pi/2$).
Using Eq.(\ref{vr}), we can express $\mathcal{M}$ as
\begin{eqnarray}
\mathcal{M}
=
c_3^2
\times
\left[
\begin{array}{cc}
c_2^2\left(1+\frac{M_1}{M_2}t_3^2\right)+\frac{M_1}{M_3}\frac{s_2^2}{c_3^2}&
-\frac{m_2}{m_1}c_2t_3\left(1-\frac{M_1}{M_2}\right)\\
-\frac{m_2}{m_1}c_2t_3\left(1-\frac{M_1}{M_2}\right)&
\frac{m_2^2}{m_1^2}\left(t_3^2+\frac{M_1}{M_2}\right)\\
-\frac{m_3}{m_1}s_2c_2\left(1+\frac{M_1}{M_2}t_3^2-\frac{M_1}{M_3}\frac{1}{c_3^2}\right)&
\frac{m_2m_3}{m_1^2}s_2t_3\left(1-\frac{M_1}{M_2}\right)\\
\end{array}
\right.
\nonumber\\
\hspace{6em}
\left.
\begin{array}{c}
-\frac{m_3}{m_1}s_2c_2\left(1+\frac{M_1}{M_2}t_3^2-\frac{M_1}{M_3}\frac{1}{c_3^2}\right)\\
\frac{m_2m_3}{m_1^2}s_2t_3\left(1-\frac{M_1}{M_2}\right)\\
\frac{m_3^2}{m_1^2}\left(s_2^2\left(1+\frac{M_1}{M_2}t_3^2\right)+\frac{M_1}{M_3}\frac{c_2^2}{c_3^2}\right)\\
\end{array}
\right]
,
\label{mass2}
\end{eqnarray}
where we define $s_2=\sin\theta^R_2$ {\it etc}.
Referring Eq.(\ref{massrel}), we can approximate $\mathcal{M}$ up to leading order as
\begin{eqnarray}
\mathcal{M}
\simeq
\left[
\begin{array}{ccc}
1&-\frac{m_2}{m_1}\theta^R_3&-\frac{m_3}{m_1}\theta^R_2\\
-\frac{m_2}{m_1}\theta^R_3&\frac{m_2^2}{m_1^2}\left((\theta^R_3)^2+\frac{M_1}{M_2}\right)&\frac{m_2m_3}{m_1^2}\theta^R_2\theta^R_3\\
-\frac{m_3}{m_1}\theta^R_2&\frac{m_2m_3}{m_1^2}\theta^R_2\theta^R_3&\frac{m_3^2}{m_1^2}\left((\theta^R_2)^2+\frac{M_1}{M_3}\right)\\
\end{array}
\right]
.
\label{mass3}
\end{eqnarray}

Requiring that Eq.(\ref{mass3}) can be diagonalized by $O=O_x(\Theta _1)\cdot O_z(\Theta _3)$ with two rotations around x- and z-axes, we can immediately diagonalize $\mathcal{M}_{\nu}$ as
\begin{eqnarray}
\mathcal{M}_{\nu}
=
\frac{m_1^2}{M_1}\times
V_{CKM}^T\left(O_{x}(\Theta _1)O_{z}(\Theta _3)\right)
\left[
\begin{array}{ccc}
\rho _1&&\\
&\rho _2&\\
&&\rho _3\\
\end{array}
\right]
\left(O_{x}(\Theta _1)O_{z}(\Theta _3)\right)^TV_{CKM}
,
\label{numass6}
\end{eqnarray}
where we define
\begin{eqnarray}
\rho _1
&=&
1+\left(\frac{\tan\Theta _3}{\cos\Theta _1}\right)x
\qquad
\rho _2
=
1-\left(\frac{\cot\Theta _3}{\cos\Theta _1}\right)x
\qquad
\rho _3
=
\frac{m_2^2}{m_1^2}\frac{M_1}{M_2}
\label{rho}
\\
\theta^R_2
&=&
-(\tan\Theta _1)\frac{m_2}{m_3}\theta^R_3
\qquad
\frac{M_2}{M_3}
=
\frac{m_2^2}{m_3^2}
\qquad
\tan 2\Theta _3
=
\frac{-2\frac{x}{\cos\Theta _1}}{\frac{x^2}{(\cos\Theta _1)^2}+\rho _3-1}
\label{theta}
\\
x
&\stackrel{def}{=}&
\frac{m_2}{m_1}\theta^R_3.
\label{xdef}
\end{eqnarray}
In what follows, we set $\Theta _1$ equals to $45^{\circ}$ as mentioned in the previous section and leave the degree of freedom of $\Theta _3$, because as seen in the following the former rotation around x-axis reduce the magnitude of $\lambda$ as $\lambda\rightarrow\lambda/\sqrt{2}$ and $\Theta _3$ cannot be taken to be maximal.

Eventually, we find the relation between CKM and MNS matrices up to $\mathcal{O}(\lambda^2)$ as follows.
\begin{eqnarray}
V_{MNS}
&=&
V_{CKM}^{\dag}\times O_{x}(45^{\circ})\times O_{z}(\Theta _3)
\nonumber\\
&\simeq&
\left[
\begin{array}{ccc}
1-\frac{\lambda^2}{2}&-\lambda&0\\
\lambda&1-\frac{\lambda^2}{2}&-A\lambda^2\\
0&A\lambda^2&1\\
\end{array}
\right]
\left[
\begin{array}{ccc}
1&0&0\\
0&\frac{1}{\sqrt{2}}&\frac{1}{\sqrt{2}}\\
0&-\frac{1}{\sqrt{2}}&\frac{1}{\sqrt{2}}\\
\end{array}
\right]
\left[
\begin{array}{ccc}
\cos\Theta _3&\sin\Theta _3&0\\
-\sin\Theta _3&\cos\Theta _3&0\\
0&0&1\\
\end{array}
\right]
\nonumber\\
&=&
\left[
\begin{array}{cc}
\left(1-\frac{\lambda^2}{2}\right)\cos\Theta _3+\frac{\lambda}{\sqrt{2}}\sin\Theta _3&
\left(1-\frac{\lambda^2}{2}\right)\sin\Theta _3-\frac{\lambda}{\sqrt{2}}\cos\Theta _3\\
\lambda\cos\Theta _3-\frac{1}{\sqrt{2}}\left(1+\left(A-\frac{1}{2}\right)\lambda^2\right)\sin\Theta _3&
\lambda\sin\Theta _3+\frac{1}{\sqrt{2}}\left(1+\left(A-\frac{1}{2}\right)\lambda^2\right)\cos\Theta _3\\
\frac{1}{\sqrt{2}}(1-A\lambda^2)\sin\Theta _3&
-\frac{1}{\sqrt{2}}(1-A\lambda^2)\cos\Theta _3\\
\end{array}
\right.
\nonumber\\
&&
\hspace{15em}
\left.
\begin{array}{c}
-\frac{\lambda}{\sqrt{2}}\\
\frac{1}{\sqrt{2}}\left(1-\left(A+\frac{1}{2}\right)\lambda^2\right)\\
\frac{1}{\sqrt{2}}(1+A\lambda^2)\\
\end{array}
\right]
\label{vrot}
\end{eqnarray}
Then, comparing with standard parametrization of $V_{MNS}$,
\begin{eqnarray*}
V_{MNS}
=
\left[
\begin{array}{ccc}
1&0&0\\
0&\cos\theta _{23}&\sin\theta _{23}\\
0&-\sin\theta _{23}&\cos\theta _{23}\\
\end{array}
\right]
\left[
\begin{array}{ccc}
\cos\theta _{13}&0&e^{-i\delta}\sin\theta _{13}\\
0&1&0\\
-e^{i\delta}\sin\theta _{13}&0&\cos\theta _{13}\\
\end{array}
\right]
\left[
\begin{array}{ccc}
\cos\theta _{12}&\sin\theta _{12}&0\\
-\sin\theta _{12}&\cos\theta _{12}&0\\
0&0&1\\
\end{array}
\right],
\end{eqnarray*}
we can immediately find the relations between observed mixing angles, i.e. $\theta _{12},\theta _{23}$ and $\theta _{13}$, and $\Theta _3,\lambda$ up to $\mathcal{O}(\lambda^2)$ :
\begin{eqnarray*}
\theta _{12}=\Theta _3-\frac{\lambda}{\sqrt{2}}
\qquad
\theta _{23}=45^{\circ}-\left(A+\frac{1}{4}\right)\lambda^2
\qquad
\theta _{13}=\frac{\lambda}{\sqrt{2}}
\qquad
\delta=\pi.
\end{eqnarray*}
Note that this model deduces the order of magnitude of $\theta _{13}$.
Though this value is not so small for $\lambda=\sin\theta _C$, it is still not conflict with the experimental data from CHOOZ experiment, i.e.
\begin{eqnarray*}
\sin^2\theta _{13}\leq 0.041\quad(\textrm{$3\sigma$ C.L.}).
\end{eqnarray*}

We can fix $\Theta _3$ using the experimental data of $\theta _{sol}(\simeq \theta _{12})$.
Combining this result with a constraint of the ratio on mass-squared differences from experimental data
\begin{eqnarray*}
\frac{\rho _2^2-\rho _1^2}{\rho _3^2-\rho _2^2}
=
\frac{\Delta m_{sol}^2}{\Delta m_{atm}^2}
\stackrel{def}{=}
\Delta
,
\end{eqnarray*}
we can finally fix the remaining dimensionless parameter, $x$ in Eqs.(\ref{rho}),(\ref{theta}), i.e.
\begin{eqnarray*}
\rho _1
=
1+\sqrt{2}(\tan\Theta _3)x
\qquad
\rho _2
=
1-\sqrt{2}(\cot\Theta _3)x
\qquad
\rho _3
=
1-2\sqrt{2}(\cot 2\Theta _3)x-2x^2
.
\end{eqnarray*}
Note that there are possibilities that $\rho _i$'s take negative values.
We, however, can always define $\rho _i$ to positive, thanks to the Majorana phases.
Using the best fit values \cite{global} :
\begin{eqnarray*}
\tan^2\theta _{sol}&=&0.39\\
\Delta m_{sol}^2=8.2\times 10^{-5}~\textrm{eV}^2
&&
|\Delta m_{atm}^2|=2.2\times 10^{-3}~\textrm{eV}^2,
\end{eqnarray*}
$\Theta _3$ is fixed as $\Theta _3=41.15^{\circ}$ and we eventually find
\begin{eqnarray*}
&&
x=-4.410
\\
&&
\rho _1=-4.450
\qquad
\rho _2=8.137
\qquad
\rho _3=-36.21
\\
&&
\left|\rho _2/\rho _1\right|=1.828
\qquad
\left|\rho _3/\rho _1\right|=8.137.
\end{eqnarray*}
Combining Eqs.(\ref{theta}),(\ref{xdef}) with these values, we get
\begin{eqnarray}
|\theta _2^R|
=
0.7329\times\sqrt{\frac{M_1}{M_3}}
\qquad
|\theta _3^R|
=
0.7329\times\sqrt{\frac{M_1}{M_2}}
.
\label{}
\end{eqnarray}
It is worthwhile noting that the relations, $\theta _{ij}^R=\mathcal{O}\left(\sqrt{M_i/M_j}\right)$, have been realized.
Note also that $\rho _3$ is larger than $\mathcal{O}(1)$.
We, however, confirm the validity of the approximation in Eq.(\ref{mass3}) since we expect that the order of magnitude of $m_1/m_2$ is similar to $m_u/m_c$ in up-quark sector from quark-lepton symmetry in $SO(10)$ GUT.
We cannot find any solutions in case of $\Delta m_{atm}^2<0$, i.e. inverted mass spectrum case.
In general, there are two possible cases reflecting the uncertainty of the sign of mass-squared difference in atmospheric neutrino oscillation experiment, normal or inverted mass spectrum, i.e. $\Delta m_{atm}^2>0$ or $\Delta m_{atm}^2<0$, respectively. 
There are some proposals to fix the sign of atmospheric neutrino mass squared difference, i.e. discrimination between normal mass spectrum and inverted one by utilizing the difference of matter effect of the earth between electron neutrino and electron anti-neutrino at Neutrino Factory\cite{Albright,Barger}.

Then, we can find absolute values of neutrino masses by using the following equation,
\begin{eqnarray*}
\mu _i
&=&
\frac{|\rho _i|}{\sqrt{\rho _2^2-\rho _1^2}}
\times
\sqrt{\Delta m_{sol}^2}.
\end{eqnarray*}
These equations lead to
\begin{eqnarray*}
\mu _1=0.5916\times 10^{-2}~\textrm{(eV)}
\qquad
\mu _2=1.082\times 10^{-2}~\textrm{(eV)}
\qquad
\mu _3=4.814\times 10^{-2}~\textrm{(eV)}.
\end{eqnarray*}

\section{The effects of CP phases and the estimation of the Majorana mass responsible for the neutrinoless double beta decay\label{three}}
In previous section, we neglect five CP phases, $\epsilon _i,\kappa _i$ and $\delta^R$.
In this section, we reanalyze neutrino mass matrix including all CP phases.
At first, embedding $\delta^R$ into $V_R$ corresponds to
\begin{eqnarray*}
V_R
&=&
\left[
\begin{array}{ccc}
\cos\theta _2^R&0&e^{-i\delta^R}\sin\theta _2^R\\
0&1&0\\
-e^{i\delta^R}\sin\theta _2^R&0&\cos\theta _2^R\\
\end{array}
\right]
\left[
\begin{array}{ccc}
\cos\theta _3^R&\sin\theta _3^R&0\\
-\sin\theta _3^R&\cos\theta _3^R&0\\
0&0&1\\
\end{array}
\right]
\\
&=&
\left[
\begin{array}{ccc}
1&&\\
&1&\\
&&e^{i\delta^R}\\
\end{array}
\right]
\left[
\begin{array}{ccc}
\cos\theta _2^R&0&\sin\theta _2^R\\
0&1&0\\
-\sin\theta _2^R&0&\cos\theta _2^R\\
\end{array}
\right]
\left[
\begin{array}{ccc}
\cos\theta _3^R&\sin\theta _3^R&0\\
-\sin\theta _3^R&\cos\theta _3^R&0\\
0&0&1\\
\end{array}
\right]
\left[
\begin{array}{ccc}
1&&\\
&1&\\
&&e^{-i\delta^R}\\
\end{array}
\right]
.
\end{eqnarray*}
This phase, however, is not a physical phase since in this paper we adopt the condition that $\theta _1^R$ equals to zero; the phase $\delta^R$ in the most left and the most right matrices in the above expression can be absorbed into $\kappa _1$ and $\kappa _2$, respectively.
Therefore, we can concentrate our attentions on four CP phases, $\epsilon _i$ and $\kappa _i$.
Next, the effect of embedding $\epsilon _i$ and $\kappa _i$ into $\mathcal{M}$ corresponds to the substitutions,
\begin{eqnarray*}
m_1\rightarrow \tilde{m}_1=e^{i\epsilon _1}m_1
\qquad
m_3\rightarrow \tilde{m}_3=e^{i	\kappa _1}m_3
\\
M_1\rightarrow \tilde{M}_1=e^{-2i\epsilon _2}M_1
\qquad
M_3\rightarrow \tilde{M}_3=e^{-2i\kappa _2}M_3.
\end{eqnarray*}
Using these substitutions, we can write down approximated expression of $\mathcal{M}=(\hat{\mathcal{M}}_D)U_R(\hat{\mathcal{M}}_R)^{-1}U_R^T(\hat{\mathcal{M}}_D)$ referring to Eq.(\ref{mass3}) as
\begin{eqnarray}
\mathcal{M}
\simeq
\frac{\tilde{m}_1^2}{\tilde{M}_1}
\times
\left[
\begin{array}{ccc}
1&-\frac{m_2}{\tilde{m}_1}\theta^R_3&-\frac{\tilde{m}_3}{\tilde{m}_1}\theta^R_2\\
-\frac{m_2}{\tilde{m}_1}\theta^R_3&\frac{m_2^2}{\tilde{m}_1^2}\left((\theta^R_3)^2+\frac{\tilde{M}_1}{M_2}\right)&\frac{m_2\tilde{m}_3}{\tilde{m}_1^2}\theta^R_2\theta^R_3\\
-\frac{\tilde{m}_3}{\tilde{m}_1}\theta^R_2&\frac{m_2\tilde{m}_3}{\tilde{m}_1^2}\theta^R_2\theta^R_3&\frac{\tilde{m}_3^2}{\tilde{m}_1^2}\left((\theta^R_2)^2+\frac{\tilde{M}_1}{\tilde{M}_3}\right)\\
\end{array}
\right]
.
\label{mass4}
\end{eqnarray}
When we diagonalize this matrix along to the same way in previous section, the conditions to satisfy $\Theta _2=0$ after rotation around x-axis ($\Theta _1$) are
\begin{eqnarray}
\theta _2^R
=
-(\tan\Theta _1)\frac{m_2}{\tilde{m}_3}\theta _3^R
\qquad
\frac{M_2}{\tilde{M}_3}
=
\frac{m_2^2}{\tilde{m}_3^2}
.
\label{cprel1}
\end{eqnarray}
To maintain the statement that $m_i,M_i$ and $\theta _i^R$ are defined by real numbers, we set $\kappa _1=-\kappa _2$ and regard $\kappa _1$ as a CP phase which embedded in mixing matrix $O_x(\Theta _1)$, i.e. $e^{\pm i\kappa _1}\sin\Theta _1$.
Thus, we can achieve correct diagonalization of $\mathcal{M}_{\nu}$ maintaining $m_i,M_i$ and $\theta _i^R$ real numbers.
This phase, $\kappa _1$, has a physical meaning as seen below.
The diagonalized mass matrix is written as
\begin{eqnarray*}
\mathcal{M}_{\nu}
=
e^{2i(\epsilon _1+\epsilon _2)}\frac{m_1^2}{M_1}
\times
V_{MNS}^{\ast}
\left[
\begin{array}{ccc}
\tilde{\rho}_1&0&0\\
0&\tilde{\rho}_2&0\\
0&0&\tilde{\rho}_3\\
\end{array}
\right]
V_{MNS}^{\dag}
,
\end{eqnarray*}
and MNS matrix can be written as
\begin{eqnarray*}
V_{MNS}
&=&
V_{CKM}^{\dag}\times O
\\
&=&
V_{CKM}^{\dag}
\left[
\begin{array}{ccc}
1&0&0\\
0&\cos\Theta _1&e^{i\kappa _1}\sin\Theta _1\\
0&-e^{-i\kappa _1}\sin\Theta _1&\cos\Theta _1\\
\end{array}
\right]
\left[
\begin{array}{ccc}
\cos\Theta _3&\sin\Theta _3&0\\
-\sin\Theta _3&\cos\Theta _3&0\\
0&0&1\\
\end{array}
\right]
\\
&=&
\left[
\begin{array}{ccc}
1&&\\
&1&\\
&&e^{-i\kappa _1}\\
\end{array}
\right]
\left[
\begin{array}{ccc}
1&&\\
&1&\\
&&e^{i\kappa _1}\\
\end{array}
\right]
V_{CKM}^{\dag}
\left[
\begin{array}{ccc}
1&&\\
&1&\\
&&e^{-i\kappa _1}\\
\end{array}
\right]
\\
&&
\times
\left[
\begin{array}{ccc}
1&0&0\\
0&\cos\Theta _1&\sin\Theta _1\\
0&-\sin\Theta _1&\cos\Theta _1\\
\end{array}
\right]
\left[
\begin{array}{ccc}
\cos\Theta _3&\sin\Theta _3&0\\
-\sin\Theta _3&\cos\Theta _3&0\\
0&0&1\\
\end{array}
\right]
\left[
\begin{array}{ccc}
1&&\\
&1&\\
&&e^{i\kappa _1}\\
\end{array}
\right]
.
\end{eqnarray*}
Rewriting
\begin{eqnarray*}
\left[
\begin{array}{ccc}
1&&\\
&1&\\
&&e^{i\kappa _1}\\
\end{array}
\right]
V_{CKM}^{\dag}
\left[
\begin{array}{ccc}
1&&\\
&1&\\
&&e^{-i\kappa _1}\\
\end{array}
\right]
=
\left[
\begin{array}{ccc}
1-\frac{\lambda^2}{2}&-\lambda&\mathcal{O}(\lambda^3)\cdot e^{-i\kappa _1}\\
\lambda&1-\frac{\lambda^2}{2}&-A\lambda^2\cdot e^{-i\kappa _1}\\
\mathcal{O}(\lambda^3)\cdot e^{i\kappa _1}&A\lambda^2\cdot e^{i\kappa _1}&1\\
\end{array}
\right],
\end{eqnarray*}
the effect of $\kappa _1$ is always suppressed by $\mathcal{O}(\lambda^2)$, so that we can follow the same procedure with Eq.(\ref{vrot}) up to $\mathcal{O}(\lambda)$.
Then, the right phase term shifts the arguments of $\tilde{\rho}_3$ as $arg(\tilde{\rho}_3)\rightarrow arg(\tilde{\rho}_3)-2\kappa _1$.
This phase has no effect on the absolute value of 3rd neutrino mass but appears in some phenomena in which observables are relevant to the Majorana phases, e.g. neutrinoless double beta decay mentioned latter in this paper.

Eventually, we can write the correct diagonal mass matrix $\mathcal{M}_{\nu}$ as
\begin{eqnarray}
\mathcal{M}_{\nu}
&=&
e^{2i(\epsilon _1+\epsilon _2)}\frac{m_1^2}{M_1}
\times
\left[
\begin{array}{ccc}
1&&\\
&1&\\
&&e^{i\kappa _1}\\
\end{array}
\right]
V_{MNS}^{\ast}
\left[
\begin{array}{ccc}
\tilde{\rho}_1&0&0\\
0&\tilde{\rho}_2&0\\
0&0&e^{-2i\kappa _1}\tilde{\rho}_3\\
\end{array}
\right]
V_{MNS}^{\dag}
\left[
\begin{array}{ccc}
1&&\\
&1&\\
&&e^{i\kappa _1}\\
\end{array}
\right]
\nonumber\\
&\stackrel{rephasing}{\longrightarrow}&
\frac{m_1^2}{M_1}
\times
V_{MNS}^{\ast}
\left[
\begin{array}{ccc}
\tilde{\rho}_1&0&0\\
0&\tilde{\rho}_2&0\\
0&0&e^{-2i\kappa _1}\tilde{\rho}_3\\
\end{array}
\right]
V_{MNS}^{\dag}
\label{numasscp1}
\\
\tilde{\rho}_1
&=&
1+\left(\frac{\tan\Theta _3}{\cos\Theta _1}\right)\tilde{x}
\label{rho1b}
\\
\tilde{\rho}_2
&=&
1-\left(\frac{\cot\Theta _3}{\cos\Theta _1}\right)\tilde{x}
\label{rho2b}
\\
\tilde{\rho}_3
&=&
1-2\left(\frac{\cot 2\Theta _3}{\cos\Theta _1}\right)\tilde{x}-\frac{1}{(\cos\Theta _1)^2}\tilde{x}^2
\label{rho3b}
\\
\tilde{x}
&\stackrel{def}{=}&
e^{-i\epsilon _1}x
\label{xdefb}
\end{eqnarray}
Note that there remain three physical parameters, $\epsilon _1,\kappa _1$ and $x$, and the phase $\epsilon _2$ does not appear explicitly, since it is not a independent parameter through $\tilde{\rho}_3$, i.e. $arg(\tilde{\rho}_3)=-2(\epsilon _1+\epsilon _2)$, as seen in Eq.(\ref{rho}). 

\begin{figure}[h]
\begin{center}
\includegraphics[scale=1.0]{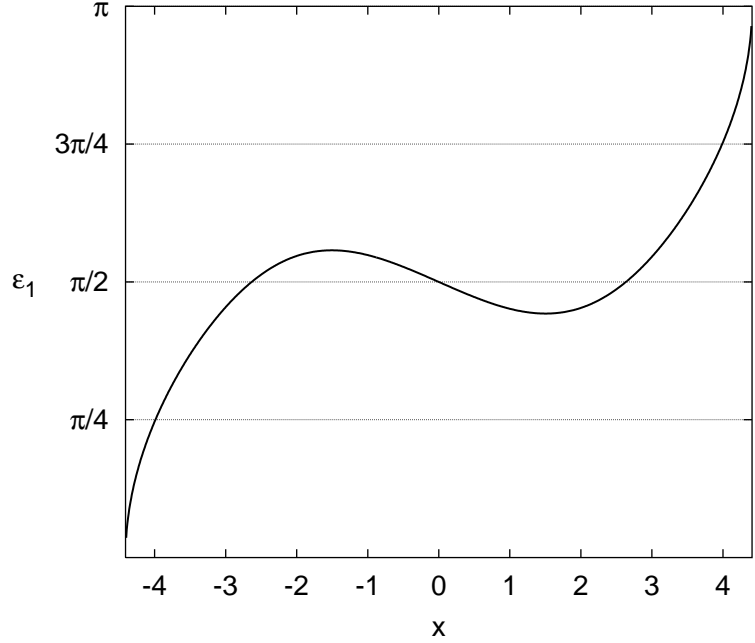}
\end{center}
\caption{Allowed values of $x$ for $\epsilon _1$ in case of $\Delta m_{atm}^2>0$.}
\label{figxphase}
\end{figure}

Furthermore, setting $\Theta _1$ equals to $45^{\circ}$, the absolute values of Eqs.(\ref{rho1b}),(\ref{rho2b}),(\ref{rho3b}) are
\begin{eqnarray}
|\tilde{\rho}_1|^2
&=&
1+2\sqrt{2}(\tan\Theta _3)(\cos\epsilon _1)x+2(\tan\Theta _3)^2x^2
\label{rho1d}
\\
|\tilde{\rho}_2|^2
&=&
1-2\sqrt{2}(\cot\Theta _3)(\cos\epsilon _1)x+2(\cot\Theta _3)^2x^2
\label{rho2d}
\\
|\tilde{\rho}_3|^2
&=&
1-4\sqrt{2}(\cot 2\Theta _3)(\cos\epsilon _1)x+4\left(2(\cot 2\Theta _3)^2-\cos 2\epsilon _1\right)x^2
\nonumber\\
&&
+8\sqrt{2}(\cot 2\Theta _3)(\cos\epsilon _1)x^3+4x^4
\label{rho3d}
\end{eqnarray}
Setting $\Theta _3=41.15^{\circ}$ and substituting these into $|\tilde{\rho}_3|^3-|\tilde{\rho}_2|^2-\Delta^{-1}(|\tilde{\rho}_2|^2-|\tilde{\rho}_2|^1)=0$, we can easily solve this equation analytically.
The result is shown in Fig.\ref{figxphase}. 

\begin{figure}[h]
\begin{center}
\includegraphics[scale=1.0]{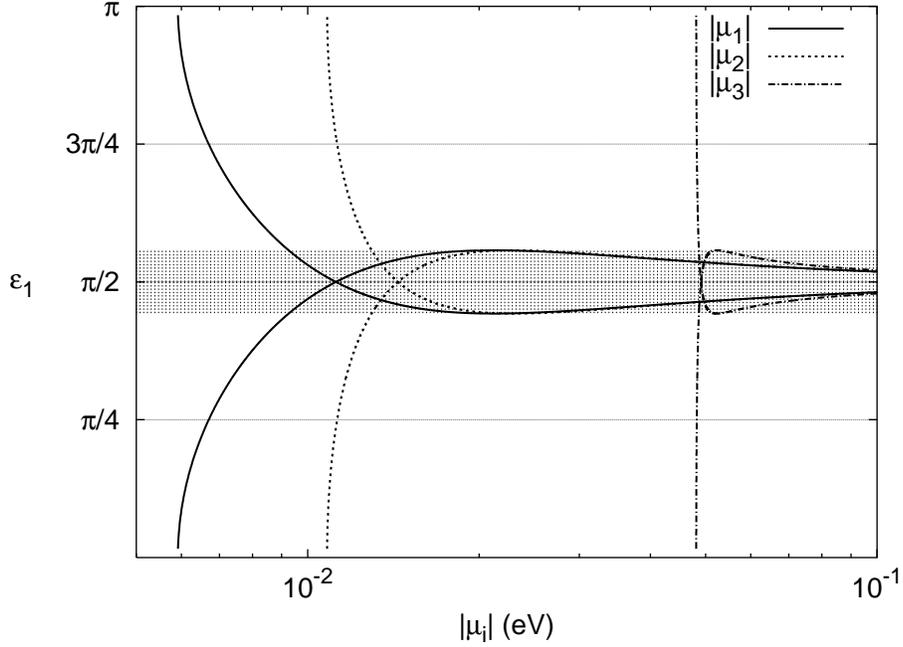}
\end{center}
\caption{
The allowed values of $|\tilde{\mu}_i|$ for $\epsilon _1$.
}
\label{figmassphase}
\end{figure}

\begin{figure}[h]
\begin{center}
\includegraphics[width=14cm,height=7cm]{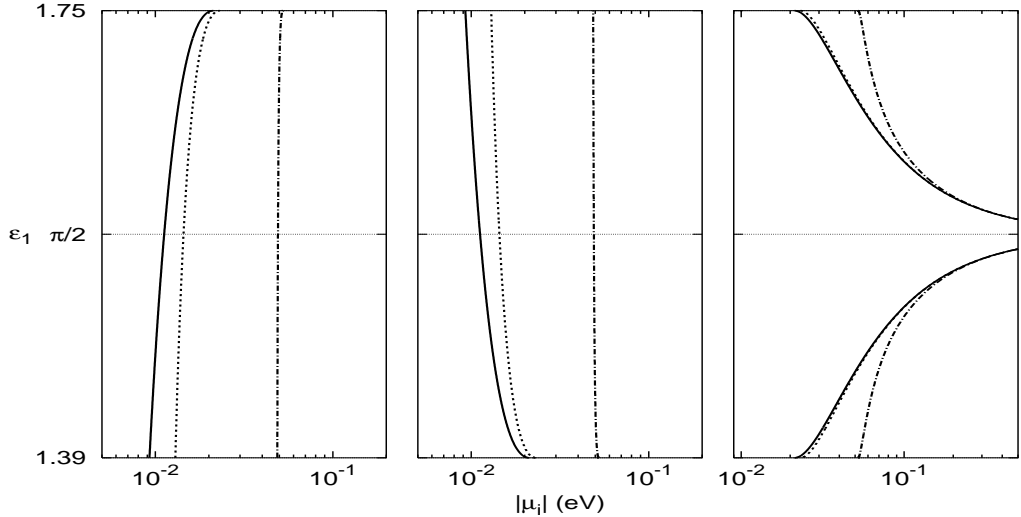}
\end{center}
\caption{
The allowed values of $|\tilde{\mu}_i|$ for $\epsilon _1$ ($1.39<\epsilon _1<1.75$).
These figures corresponds to the shaded area in Fig.\ref{figmassphase}.
}
\label{figmassphase3}
\end{figure}

Using above analytical solution, we can also estimate the absolute values of neutrino masses, $|\tilde{\mu}_1|,|\tilde{\mu}_2|$ and $|\tilde{\mu}_3|$ by using the following equation,
\begin{eqnarray}
|\tilde{\mu}_i|
&=&
\frac{|\tilde{\rho}_i|}{\sqrt{|\tilde{\rho}_2|^2-|\tilde{\rho}_1|^2}}
\times\sqrt{\Delta m_{sol}^2},
\label{numassabs}
\end{eqnarray}
and the obtained results are shown in Fig.\ref{figmassphase}.
We find that there exist allowed values of $|\tilde{\mu}_i|$ for any values of $\epsilon _1$ in case of $\Delta m_{atm}^2>0$ while we cannot find any allowed values of $|\tilde{\mu}_i|$ in case of $\Delta m_{atm}^2<0$.
We can find that the result shown in Fig.\ref{figmassphase} has complicated structure in certain region of $\epsilon _1$, i.e. $1.39<\epsilon _1<1.75$ (the shaded area corresponds to this region).
This is because as seen in Fig.\ref{figxphase} there are two or three solutions of $x$ for $\epsilon _1$, there are also multi solutions of $|\tilde{\mu}_i|$ for $\epsilon _1$ in this region.
We discriminate these solutions to three parts in Fig.\ref{figmassphase3}.

\begin{figure}[h]
\begin{center}
\includegraphics[scale=0.9]{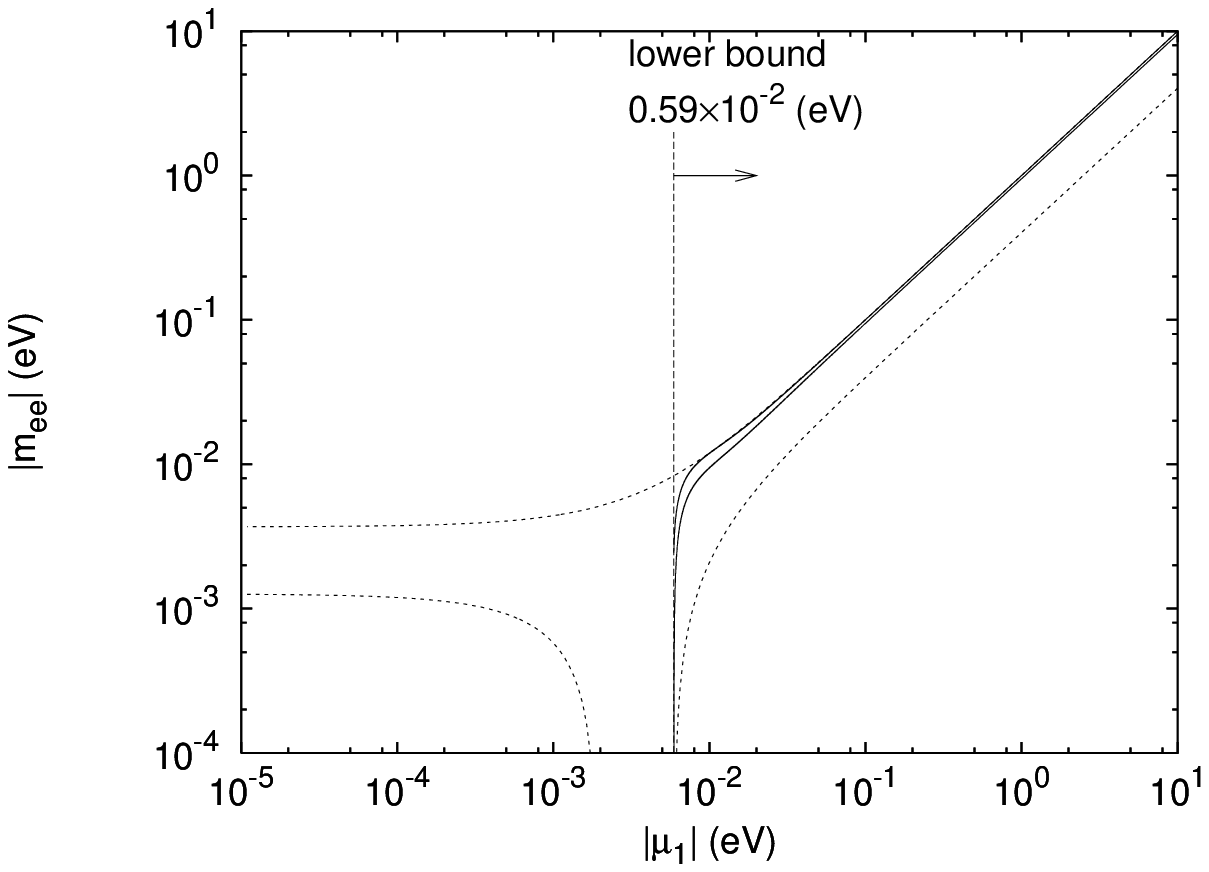}
\end{center}
\caption{
Majorana mass $|m_{ee}|$ in case of $\Delta m_{atm}^2>0$.
The dotted curve lines correspond to the general two independent Majorana phases case and solid curve lines correspond to the numerical results of our model.
Note that in our model there exists a lower limit of $|\tilde{\mu}_1|$, i.e. $|\tilde{\mu}_1|\geq 0.59\times 10^{-2}$ eV (the vertical dashed line), as seen in Fig.\ref{figmassphase}.
}
\label{figmajomass}
\end{figure}

\begin{figure}[h]
\begin{minipage}[t]{0.45\linewidth}
\includegraphics[width=\linewidth]{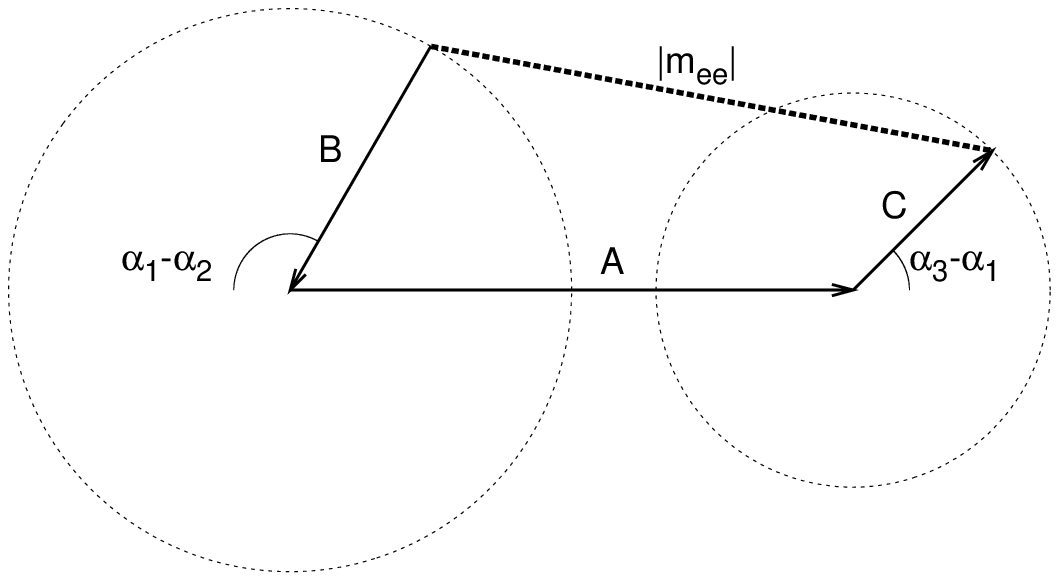}
\caption{A schematic view of Majorana mass.}
\label{figschmajo}
\end{minipage}
\hspace{5mm}
\begin{minipage}[t]{0.45\linewidth}
\includegraphics[width=\linewidth]{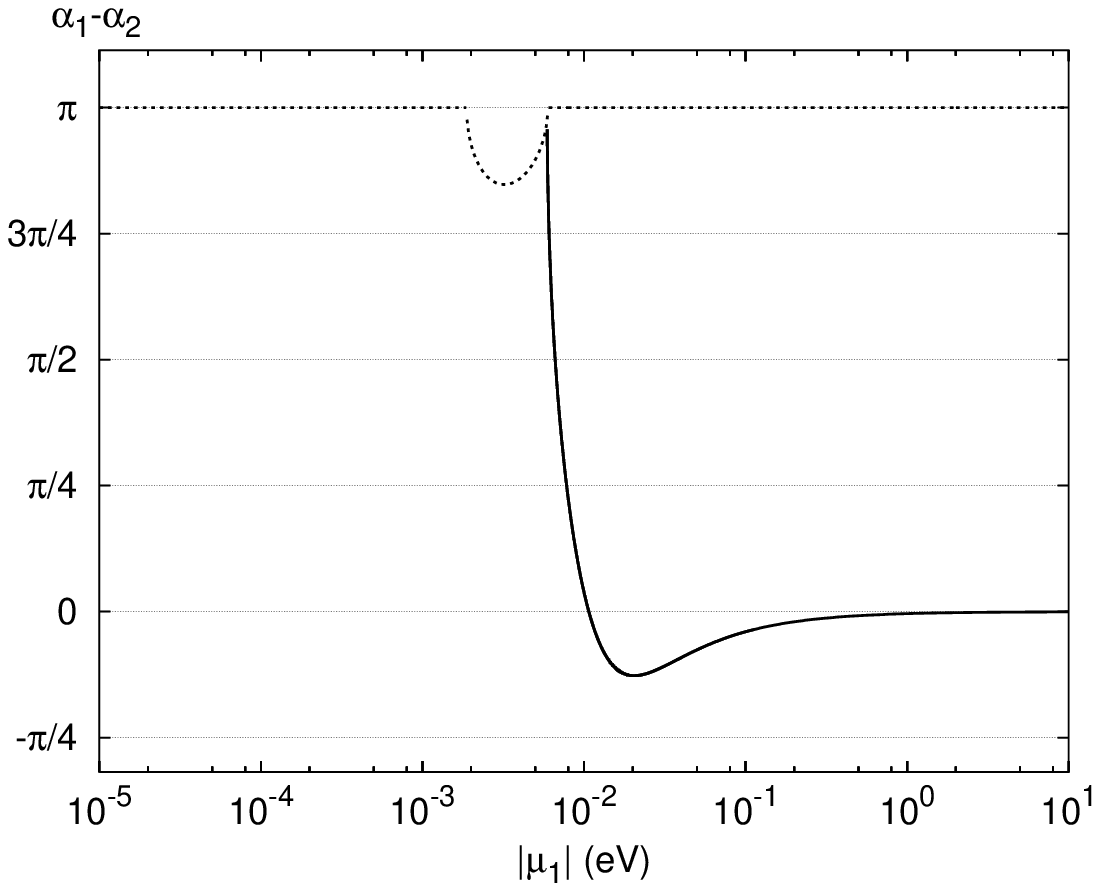}
\caption{
Maximum values of $\alpha _1-\alpha _2$ for $|\tilde{\mu}_1|$.
The dotted line corresponds to the general two independent Majorana phases case and solid line corresponds to our model.
}
\label{figmaxphase}
\end{minipage}
\end{figure}

Next, we deduce the Majorana mass responsible for the neutrinoless double beta decay experiments \cite{Joaquim}.
Defining $arg(\tilde{\rho}_i)=\alpha _i$ and Using Eq.(\ref{numasscp1}), we can write $\mathcal{M}_{\nu}$ as
\begin{eqnarray}
\mathcal{M}_{\nu}
&\propto&
V_{MNS}^{\ast}
\left[
\begin{array}{ccc}
e^{i\alpha _1}|\tilde{\mu}_1|&0&0\\
0&e^{i\alpha _2}|\tilde{\mu}_2|&0\\
0&0&e^{i\alpha _3-2i\kappa _1}|\tilde{\mu}_3|\\
\end{array}
\right]
V_{MNS}^{\dag}
=
U^{\ast}
\left[
\begin{array}{ccc}
|\tilde{\mu}_1|&0&0\\
0&|\tilde{\mu}_2|&0\\
0&0&|\tilde{\mu}_3|\\
\end{array}
\right]
U^{\dag}
\nonumber\\
U
&=&
V_{MNS}
\cdot
diag\left(1,e^{\frac{i}{2}(\alpha _1-\alpha _2)},e^{\frac{i}{2}(\alpha _1-\alpha _3+2\kappa _1)}\right)
,
\label{majocp}
\end{eqnarray}
where we neglect the overall phase.
The two remaining phases, $\alpha _1-\alpha _2$ and $\alpha _1-\alpha _3+2\kappa _1$, are known as {\it "Majorana phases"}.
The definition of Majorana mass responsible for the neutrinoless double beta decay experiments is
\begin{eqnarray}
|m_{ee}|
&=&
\left|\sum _{i=1}^3~|\tilde{\mu}_i|(U^{\ast}_{1i})^2\right|
\nonumber\\
&=&
\left|
|\tilde{\mu}_1|\cos^2\theta _{12}\cos^2\theta _{13}
+
|\tilde{\mu}_2|\sin^2\theta _{12}\cos^2\theta _{13}e^{i(\alpha _2-\alpha _1)}
+
|\tilde{\mu}_3|\sin^2\theta _{13}e^{i(\alpha _3-\alpha _1-2\kappa _1)}
\right|
.
\nonumber\\
\label{majomass}
\end{eqnarray}
In general, this quantity has two independent CP phases and especially for the case of normal mass spectrum, $|m_{ee}|$ can take zero accidentally in certain region of $|\tilde{\mu}_1|$, $1.88\times 10^{-3}\textrm{ eV}<|\tilde{\mu}_1|<5.97\times 10^{-3}\textrm{ eV}$.
We, however, already found in Fig.\ref{figxphase} that the phase $\epsilon _1$ depends on $x$ in this model and this means that one Majorana phase, $\alpha _1-\alpha _2$, is a function of $|\tilde{\mu}_1|$ shown in Fig.\ref{figmaxphase}.
On the other hand, the other Majorana phase, $\alpha _3-\alpha _1-2\kappa _1$, is completely independent phase reflecting the uncertainly of $\kappa _1$.
Therefore, we expect that the uncertainly of $|m_{ee}|$ by these two phases are strongly suppressed.
The numerical results of $|m_{ee}|$ are shown in Fig.\ref{figmajomass}.
We find in this figure that the allowed region of $|m_{ee}|$ is suppressed considerably compared with the general case.

This figure can be understood as follows.
In general, $|m_{ee}|$ can be written
\begin{eqnarray*}
|m_{ee}|
&=&
\left|
A+Be^{i(\alpha _2-\alpha _1)}+Ce^{i(\alpha _3-\alpha _1)}
\right|
=
\left|
A-Be^{i(\alpha _2-\alpha _1+\pi)}+Ce^{i(\alpha _3-\alpha _1)}
\right|
\\
A
&=&
|\tilde{\mu}_1|\times\cos^2\theta _{12}\cos^2\theta _{13}
\\
B
&=&
\sqrt{|\tilde{\mu}_1|^2+\Delta m_{sol}^2}\times\sin^2\theta _{12}\cos^2\theta _{13}
\\
C
&=&
\sqrt{|\tilde{\mu}_1|^2+\Delta m_{sol}^2+\Delta m_{atm}^2}\times\sin^2\theta _{13}
,
\end{eqnarray*}
and the schematic view of the relation between these quantities are shown in Fig.\ref{figschmajo}.
The necessary condition to minimize $|m_{ee}|$ is
\begin{eqnarray}
\alpha _1-\alpha _2
&=&
\arccos\left(\frac{C^2-A^2-B^2}{2AB}\right)
\qquad(1.88\times 10^{-3}\textrm{eV}\leq|\tilde{\mu}_1|\leq 5.97\times 10^{-3}\textrm{eV})
\nonumber
\\
&=&
\pi
\qquad(\textrm{else}).
\label{maxphase}
\end{eqnarray}
We, however, can find in Fig.\ref{figmaxphase} that the maximal values of $\alpha _1-\alpha _2$ are significantly smaller than the general case in most of the region.

\section{Conclusion\label{concl}}
In this paper, we discuss the extra rotations induced by additional diagonalization of Majorana mass matrix and derive the absolute values of three neutrino masses and Majorana mass responsible for the neutrinoless double beta decay experiment only invoking to the seesaw mechanism collaborated by the unification of neutrino Dirac mass matrix with that of up-type quarks and the left-right symmetry based on $SO(10)$ GUT.

We specify these extra rotations to bi-maximal rotations around x- and z-axes and find that these extra rotations can explain the interesting and nontrivial relations between CKM and MNS matrices.
In this analysis, we find the specific value of $\theta _{13}$, which does not conflict with the experimental data at $3\sigma$ C.L.

In Sec.\ref{two}, we ignore CP phases for simplicity and find that the absolute values of neutrino masses satisfy $m_1:m_2:m_3\approx 1:2:8$, i.e. neutrino masses have hierarchical structure, and that there is no solution in the case of inverted mass spectrum.

In Sec.\ref{three}, we reanalyze the absolute values of neutrino masses and Majorana mass responsible for the neutrinoless double beta decay experiment, by including all CP phases.
In this analyses, we find that only two CP phases, $\epsilon _1$ and $\kappa _1$, remain as independent degrees of freedom.
The former phase has a physical meaning both in the analyses of the absolute values of neutrino masses and Majorana mass responsible for the neutrinoless double beta decay experiment, while the latter phase appears only in the analysis of Majorana mass responsible for the neutrinoless double beta decay experiment.
In the analysis of the absolute values of neutrino masses, we cannot decide them uniquely but find these quantities have well-defined lower bounds though we cannot find any solutions in the case of inverted mass spectrum.
In the analysis of Majorana mass responsible for the neutrinoless double beta decay experiment, we find that one Majorana phase is a function of $|\tilde{\mu}_1|$ and that this reduces the allowed region of $|m_{ee}|$ considerably.
\begin{acknowledgments}
I thank C. S. Lim for helpful and fruitful discussions and for careful reading and correcting of the manuscript.
\end{acknowledgments}

\end{document}